\documentclass[twocolumn]{aastex631}
\usepackage[T1]{fontenc}
\usepackage{newtxtext, newtxmath}
\usepackage{natbib}
\usepackage{graphicx}
\usepackage{amsmath}
\usepackage{amssymb}
\graphicspath{{figures/}}
\newcommand{\vice}{\textsc{vice}}
\newcommand{\heistar}{He \textsc{i}$^*$}
\newcommand{\heratio}{\ensuremath{
    ^3\text{He}/^4\text{He}
}}
\newcommand{\msun}{\ensuremath{\text{M}_\odot}}
\newcommand{\hii}{H \textsc{ii}}
\newcommand{\hi}{H \textsc{i}}

\newcommand{\ddfrac}[2]{\ensuremath{\frac{
    \displaystyle{#1}
}{
    \displaystyle{#2}}
}}

\newcommand{\carnegieaffil}{%
Carnegie Science Observatories, 813 Santa Barbara St., Pasadena, CA 91101, USA}
\newcommand{\durhamcentreaffil}{%
Centre for Extragalactic Astronomy, Durham University, Durham, DH1 3LE, UK}
\newcommand{\durhamdeptaffil}{%
Department of Physics, Durham University, South Road, Durham DH1 3LE, UK}
\newcommand{\osuaffil}{%
Department of Astronomy, The Ohio State University, 140 W. 18th Ave., Columbus, OH 43210, USA}
\newcommand{\ccappaffil}{%
Center for Cosmology \& AstroParticle Physics (CCAPP), The Ohio State University, 191 W. Woodruff Ave., Columbus, OH 43210, USA}

\begin{document}

% \title{On the Chemical Evolution of Hydrogen and Helium Isotopes}
% \shorttitle{Chemical Evolution of H and He Isotopes}

% \title{The Constraining Power of the \heratio\ Ratio for Chemical Evolution in the Milky Way}
% \shorttitle{Evolution of the \heratio\ Ratio}

% \title{Metals versus Non-metals: Chemical Evolution of Hydrogen and Helium Isotopes in the Milky Way}
% \shorttitle{Chemical Evolution of H and He Isotopes}

% \title{Metals versus Non-metals: A Case for Abundance Measurements of Hydrogen and Helium Isotopes in the Nebular Phase}
% \shorttitle{The Constraining Power of H and He Isotopes}

\title{Metals versus Non-metals: Chemical Evolution of Hydrogen and Helium Isotopes in the Milky Way}
\shorttitle{Metals versus Non-metals}

\author[0000-0002-6534-8783]{James W. Johnson}
\affiliation{\carnegieaffil}

% \author{Co-Conspirators}
% \shortauthors{J.W. Johnson et al.}

\author[0000-0003-4912-5157]{Miqaela K. Weller}
\affiliation{\osuaffil}
\affiliation{\ccappaffil}

\author[0000-0001-7653-5827]{Ryan J. Cooke}
\affiliation{\durhamcentreaffil}
\affiliation{\durhamdeptaffil}

\shortauthors{J.W. Johnson, M.K. Weller \& R.J. Cooke}

\begin{abstract}
\noindent
Star formation drives changes in the compositions of galaxies, fusing H and He into heavier nuclei.
This paper investigates the differences in abundance evolution between metal and non-metal isotopes using recent models of Galactic chemical evolution appropriate for the thin disk epoch.
A strong degeneracy arises between metal yields from stellar populations and the mean Galactocentric radial velocity of the interstellar medium (ISM).
Similar metallicities arise when increases (decreases) in metal yields are combined with increases (decreases) to the gas flow velocity.
A similar degeneracy exists between metal yields and the rate of gas ejection from the ISM.
We demonstrate that this degeneracy can be confidently broken with precise measurements of the hydrogen (D/H) and helium (\heratio) isotope ratios in the Galactic ISM.
At fixed O/H, higher metal yields lead to higher D/H and lower \heratio.
Measurements available to date are not sufficiently precise or numerous to draw confident conclusions.
A detailed inventory of non-metal isotopes in the Milky Way would provide critical empirical constraints for stellar and galactic astrophysics, as well as a new test of Big Bang Nucleosynthesis.
We forecast that only $\sim$4 additional measurements of \heratio\ within $\sim$$3$ kpc of the Sun are required to measure the primordial \heratio\ ratio at $\sim$30\% precision.
In parallel, empirical benchmarks on metal yields also have the power to inform stellar models, since absolute yield calculations carry factor of $\sim$$2-3$ uncertainties related to various complex processes (e.g., rotational mixing, convection, mass loss, failed supernovae).
\end{abstract}

\section{Introduction}
\label{sec:intro}

The chemical compositions of gas and stars encode a wealth of information on the assembly and evolution of their host galaxies.
% Modern spectroscopic surveys have measured and are continuing to measure metal abundances in the Milky Way (e.g., Milky Way Mapper: {\color{red} Kollmeier et al. 2025}; Gaia-ESO: {\color{red} Name et al. 20XX}) and external galaxies (e.g., MaNGA: {\color{red} Bundy et al. 2015}; 
The information in these measurements is often interpreted through the lens of Galactic chemical evolution (GCE) models (see discussion in, e.g., the reviews by \citealt{Tinsley1980} and \citealt{Matteucci2021}).
However, these models are affected by uncertainties in key quantities.
In particular, the total mass of metals produced by stellar populations is difficult to constrain empirically (see discussion in, e.g., \citealt{Weinberg2024}).
This metal production factor, or ``yield,'' is degenerate with the amount of mass that gets ejected from the interstellar medium (ISM; e.g., \citealt{Sandford2024}).
In this paper, we show that GCE models incorporating radial gas flows \citep[e.g.,][]{Lacey1985, Bilitewski2012} are affected by a similar degeneracy.
We further demonstrate that abundance measurements of non-metal isotopes in the interstellar medium would not only break this degeneracy, but also provide useful constraints for stellar evolution and a new test of current models of Big Bang Nucleosynthesis (BBN; see, e.g., the seminal works by \citealt{Alpher1948, Hoyle1964, Peebles1966, Wagoner1967}, and the more recent reviews by \citealt{Schramm1998, Steigman2007, Cooke2024}).
\par
Stellar models do not confidently predict metal yields.
Poorly understood processes, such as mass loss \citep[e.g.,][]{Sukhbold2016}, convection and convective boundaries \citep[e.g.,][]{Chieffi2001, Ventura2013}, and nuclear reaction rates \citep[e.g.,][]{Herwig2004, Herwig2005, Kotar2025}, introduce uncertainties of a factor of a $\sim$few in the mass of metals released to the ISM at the end of a star's life (see discussion in, e.g., \citealt{Johnson2023b} in the context of nitrogen).
This paper highlights prior work demonstrating similar effects due to rotation \citep{Limongi2018} and black hole formation \citep{Griffith2021} in high mass stars (see discussion in Section \ref{sec:discussion:yield-scale} below).
This uncertainty is problematic for GCE models, since stellar yields play a direct role in establishing metal abundances.
Previous GCE models with sufficiently large yields often incorporate ejection, which lowers metallicity by removing metals from the ISM and replacing them with hydrogen gained through accretion (see discussion in, e.g., \citealt{Weinberg2017b}).
This connection results in a classic ``source-sink'' degeneracy, whereby metals can be more efficiently produced but also more efficiently ejected, resulting in similar present-day abundances \citep[e.g.,][]{Hartwick1976, Cooke2022, Johnson2023c, Sandford2024}.
% A reliable understanding of the scale of metal production by stars would break this degeneracy.
\par
Stellar yields are also difficult to constrain empirically.
One method, perhaps the most direct, would be to simply count up all of the metals and all of the stars in the local Universe (e.g., \citealt{Prochaska2003, Gallazzi2008, Peeples2014}; see also the review by \citealt{Peroux2020}).
However, this approach requires reliable measurements of the metallicities of circumgalactic media \citep[CGM;][]{Tumlinson2017}.
The CGM metallicity varies substantially both between and within individual halos, making it difficult to pin down mass-weighted averages \citep[e.g.,][]{Zahedy2019, Zahedy2021, Cooper2021, Haislmaier2021, Kumar2024, Sameer2024}.
Overall metal abundances are also subject to uncertainties in abundance determination (see, e.g., the review by \citealt{Kewley2019}).
\par
A second diagnostic of the scale of stellar yields comes from observing supernovae (SNe) themselves.
\citet{Rodriguez2023} measured the mean Fe yield of Type II SNe by examining the radioactive tails of their lightcurves.
\citet{Weinberg2024} extended their measurement to infer population-averaged yields, which quantify the metal mass produced per unit mass of star formation.
In addition to the statistical uncertainty itself, their inference is also affected by systematic uncertainties in the number of SNe that arise from a stellar population of a given mass \citep[e.g.,][]{Pejcha2015, Ertl2016, Sukhbold2016}.
The ratio of $\alpha$-capture and iron-peak element yields from massive stars is also an important source of uncertainty with this method (see discussion in \citealt{Sit2025}).
\par
A third diagnostic of stellar yields is rooted in trends in metallicity with stellar age.
The efficient ejection, coupled with efficient metal production to maintain abundances at plausible levels, has a secondary effect on enrichment timescales, which in turn influences age-metallicity trends.
In \citet{Johnson2025}, we focused on the recent empirical result that stellar metallicities exhibit a remarkably flat trend with age \citep[e.g.,][]{Spina2022, daSilva2023, Willett2023, Gallart2024}.
Across much of the Galactic disk, metallicity declines by only $\sim$$0.1$ dex between ages of $\tau \sim 0$ and $\sim$$10$ Gyr \citep{Roberts2025}.
Our models in \citet{Johnson2025} reproduced this flat trend more readily if stars make $\sim$twice as many metals as inferred by \citet{Weinberg2024}.
The required level of metal production is smaller if accreting gas is pre-enriched to low metallicity (see discussion in \citealt{Johnson2025}) or if the rates of all SNe are higher at low metallicity \citep[e.g.,][]{Gandhi2022, Pessi2023, Johnson2023a}.
Nuance regarding these processes makes it challenging to pin down the scale of yields using stellar ages, which are notoriously difficult to measure as well \citep[e.g.,][]{Soderblom2010, Chaplin2013}.
\par
Each of these diagnostics of metal production are affected by their own sources of systematic uncertainty.
In light of these challenges, this paper advocates that the community dedicate effort to an additional approach.
In particular, we show that this degeneracy between stellar yields, the rate of ejection, and the radial gas velocity can be broken using precise measurements of the isotope ratios of non-metals.
The data products would not only break this fundamental degeneracy in GCE models, but also enable the first estimate of the primordial helium isotope ratio, $(\heratio)_p$.
\par
This paper is organized as follows.
We describe \heratio\ and D/H measurements presently available in the literature in Section \ref{sec:data} below.
We describe our GCE models in Section \ref{sec:gce}.
We illustrate the yield-ejection-velocity degeneracy and demonstrate that it can be broken with D/H and \heratio\ in Section \ref{sec:results}.
We discuss techniques for measuring non-metal isotope ratios and the potential impact of the data products in Section \ref{sec:discussion}.
We conclude in Section \ref{sec:conclusions}.

\section{Data}
\label{sec:data}

We use measurements of \heratio\ and D/H available in the literature.
\citet{Mahaffy1998} measured both isotope ratios in Jupiter's atmosphere using data from the Galileo Probe Mass Spectrometer \citep{Niemann1992}.
These measurements can be attributed to the composition of the protosolar nebula.
We also use measurements of D/H in the local ISM compiled by \citet{Linsky2006}, which are drawn from prior work available at the time (see references therein).
\par
Additional measurements of \heratio\ in the Milky Way (MW) have been reported by \citet{Balser2018} and \citet{Cooke2022}.
\citet{Cooke2022} measured \heratio\ along the line of sight toward a hot massive star in the Orion Nebula, $\Theta^2$A Ori \citep{O'Dell1993}, using a combination of optical and infrared transitions ($\lambda 3188$ \AA, $\lambda 3889$ \AA, and $\lambda 1.0833\ \mu$m) of metastable helium, \heistar.
\citet{Balser2018} measured $^3$He$^+$/H$^+$ and $^4$He$^+$/H$^+$ in five Galactic \hii\ regions using spin-flip transitions and radio recombination lines in the $8-10$ GHz frequency range using the Green Bank Telescope.
We omit one of these \hii\ regions (Sh 2-209) because its distance is unclear.
The \citet{Balser2018} catalog lists a Galactocentric radius of $R = 16.2$ kpc, but \citet{Yasui2023} found $R = 10.3$ kpc based on near-infrared imaging and astrometric data from Gaia EDR3 \citep{GaiaCollaboration2021}.
The helium isotope ratio of Sh 2-209 is $\heratio = (1.41 \pm 0.28) \times 10^{-4}$, which is consistent with our adopted primordial ratio (see Section \ref{sec:gce:yields:helium} below).
A Galactocentric radius of $R = 10.3$ kpc would therefore place this \hii\ region slightly but not significantly below the rest of the \citet{Balser2018} sample (see Figure \ref{fig:present-day-profiles} below).

\section{Galactic Chemical Evolution Models}
\label{sec:gce}

We use multi-zone GCE models from \citet{Johnson2025} and \citet{Johnson2025-solo}.
Specifically, we use the parameter choices that are broadly consistent with the equilibrium scenario for the MW.
We integrate these models numerically using the publicly available \textsc{Versatile Integrator for Chemical Evolution} (\vice; \citealt{Johnson2020})\footnote{
    Install: \url{https://pypi.org/project/vice}.
    Documentation: \url{https://vice-astro.readthedocs.io}. Source Code: \url{https://github.com/giganano/VICE.git}.
}.
In this class of GCE models, the ISM metallicity rapidly approaches some steady-state, after which increases in the overall abundance are negligible (see also \citealt{Larson1972} and \citealt{Weinberg2017b}).
\citet{Johnson2025} demonstrated that this type of enrichment history reproduces trends in metallicity with stellar population age much more accurately than models predicting larger increases in abundances during the thin disk epoch.
We adopt our models directly from \citet{Johnson2025} and \citet{Johnson2025-solo}, making no modifications with the exception of including helium and deuterium in the \vice\ integrations.
\par
These models discretize the Galactic disk into $\delta R = 100$ pc rings between $R = 0$ and $20$ kpc, capturing the effects of enrichment in different regions of the MW.
Each annulus is coupled to its nearest neighbors through the exchange of gas and stellar populations, but enrichment is otherwise described by a conventional ``one-zone'' GCE model where stellar yields mix instantaneously and homogeneously with the star forming ISM (see, e.g., the reviews by \citealt{Tinsley1980} and \citealt{Matteucci2021}).
We determine the present-day Galactocentric radius of each stellar population by sampling from a normal distribution centered on the radius of formation.
The width of the distribution, and by extension the distances stars have migrated, increases with population age.
The Galaxy follows the observed Kennicutt-Schmidt relation \citep{Schmidt1959, Schmidt1963, Kennicutt1998}, wherein the surface densities of gas and star formation follow a single power-law relation, $\dot\Sigma_\star \propto \Sigma_g^N$, with $N = 1.5$.
\par
Each annulus follows a star formation history (SFH) that initially rises rapidly before falling for the remainder of the disk lifetime, according to
\begin{equation}
\dot\Sigma_\star \propto
\left(1 - e^{-t / \tau_\text{rise}}\right)
e^{-t / \tau_\text{sfh}}.
\end{equation}
In each of our models, the values of $\tau_\text{rise}$ and $\tau_\text{sfh}$ increase with Galactocentric radius such that the models reproduce the observed profile in stellar age.
The exact prescription for $\tau_\text{rise}$ and $\tau_\text{sfh}$ differs slightly between our ejection- and flow-driven models, but the differences are unimportant to this paper.
The radial flow models from \citet{Johnson2025-solo} use pure exponential scalings in $\tau_\text{rise}$ and $\tau_\text{sfh}$ with scale lengths of $r_\text{rise} = 6.5$ kpc and $r_\text{sfh} = 4.7$ kpc, respectively.
Our ejection-driven models from \citet{Johnson2025} use a piece-wise prescription that nonetheless closely resembles this simple exponential.
These prescriptions encode the ``inside-out'' nature of disk formation, wherein the inner regions assemble first on short timescales followed by the outer regions on longer timescales (see discussion in, e.g., \citealt{Bird2013}).
% Each stellar population follows a \citet{Kroupa2001} initial mass function (IMF).

% \begin{itemize}

%     \item We use multi-zone GCE models from \citet{Johnson2025} and {\color{red} Johnson (2025, in preparation)}, which we integrate numerically using the \textsc{Versatile Integrator for Chemical Evolution} (\vice; \citealt{Johnson2020}).
%     These models discretize the Galactic disk into $\delta R = 100$ pc rings in order to model enrichment in different regions of the MW simultaneously.

%     \item We use the models from these papers that are broadly consistent with the equilibrium scenario.
%     {\color{red} Brief summary of equilibrium scenario}.

%     \item Radial migration follows the same prescription as in previous work.
%     {\color{red} Brief summary}, each final radius is sampled from a Gaussian centered on the birth radius.

%     \item Star formation efficiency follows the same prescription as in previous work, too.
%     {\color{red} Brief summary}, the disk follows the Kennicutt-Schmidt relation.

% \end{itemize}

\subsection{Stellar Yields}
\label{sec:gce:yields}

Motivated by uncertainties in the overall scale of stellar yields (see discussion in Section \ref{sec:intro}), we follow two assumed normalizations in this paper.
The first follows the recommendation by \citet{Weinberg2024} based on the determination of the mean Fe yield from the radioactive tails of Type II SN lightcurves by \citet{Rodriguez2023}.
The mass fraction of a given stellar population's initial mass later released to the ISM in a given element is approximately equal to its solar abundance by mass.
We refer to this overall yield scale as $y / Z_\odot = 1$.
Our second normalization is a factor of two larger, so we refer to this assumption as $y / Z_\odot = 2$.
\par
The yields of D, $^3$He, and $^4$He are significantly more well-understood than metal production.
We discuss how we accommodate non-metals at the different scales individually.
$^3$He and $^4$He have significant contributions from asymptotic giant branch (AGB) stars.
% \citet{Johnson2025} and \red{Johnson (2025)} do not discuss AGB star nucleosynthesis.
\vice\ handles this enrichment channel by determining the total mass of dying stars from all timesteps leading up to the current time according to
\begin{equation}
\dot \Sigma_x^\text{AGB} = \int_0^t
y_x^\text{AGB}(M_\text{TO}, Z)
\dot \Sigma_\star(t') \dot h(t - t') dt',
\end{equation}
where $h$ describes the fraction of a single stellar population's mass that is in the form of main sequence stars at a given age.
\vice\ determines $h(t)$ based on a specified initial mass function, for which we use the \citet{Kroupa2001} form in this work, and a stellar mass-lifetime relation, for which we use the \citet{Larson1974} form.
The yield, $y_x^\text{AGB}$, is evaluated at the main sequence turnoff mass, $M_\text{TO}$, and metallicity of stellar populations at each previous timestep, integrated up to the current time, $t$.
Further details on the implementation of AGB star yields in \vice\ can be found in \citet{Johnson2020} and \citet{Johnson2023b}.

% \begin{itemize}

%     \item Motivated by uncertainties in the overall scale of stellar yields (see discussion in Section \ref{sec:intro}), we follow two assumed scales in this paper.
%     The first follows the recommendation by {\color{red} Weinberg et al. (2024)} based on the radioactive tails of Type II SNe lightcurves {\color{red} (Rodriguez et al. 2023)}, which is that the mass fraction of a given stellar population converted into some element $x$ is approximately equal to its solar abundance by mass.
%     For example, a hypothetical $1000$ \msun\ stellar population would produce $5.72$ \msun of O based on a solar O abundance of $Z_{\text{O},\odot} = 0.00572$ {\color{red} (Asplund et al. 2009)}.
%     We refer to this overall yield scale as $y / Z_\odot = 1$.
%     Our second normalization elevates the O yield from this baseline value by a factor of $\sim$2, so we refer to this assumption as $y / Z_\odot = 2$.

%     \item The yields of D and He are more well understood than metal yields.
%     We discuss how our models accommodate non-metals at different yield scales below.

% \end{itemize}

\subsubsection{Oxygen}
\label{sec:gce:yields:oxygen}

\begin{figure}
\centering
\includegraphics[scale = 1]{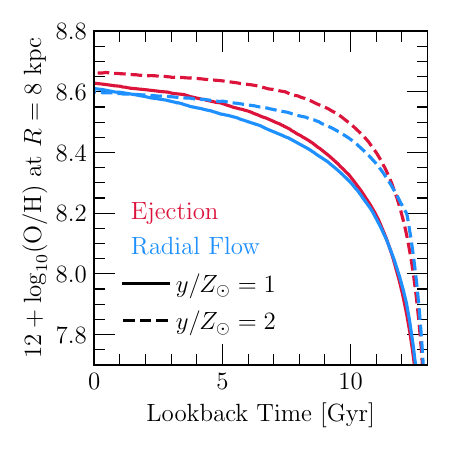}
\caption{
Evolution of the ISM O abundance with time at $R = 8$ kpc in our four GCE models.
Models using ejection of ISM gas from the disk versus those using radial flows within the disk (see discussion in Section \ref{sec:gce:ejection-vs-radialflows}) are color coded red and blue, respectively.
Models using the $y / Z_\odot = 1$ scale of stellar yields versus those using $y / Z_\odot = 2$ (see discussion in Section \ref{sec:gce:yields}) are marked as solid and dashed lines, respectively.
\textbf{Summary}: Models enrich on different timescales depending on the scale of metal yields but otherwise lead to the same O/H abundances in the ISM at the present day.
}
\label{fig:degeneracy}
\end{figure}

We focus on O as the representative metal in this paper.
Stellar populations release oxygen on timescales much shorter than the dynamical time of the Galactic disk, and O is commonly measured in the gas-phase, making it a useful diagnostic for both theoretical and observational purposes.
Oxygen makes up 0.572\% of the mass of the Sun based on the photospheric abundance measurements by \citet{Asplund2009}.
A hypothetical $1000$ \msun\ stellar population would therefore produce $5.72$ \msun\ of O under the $y / Z_\odot = 1$ normalization.
Newly produced O appears in the ISM immediately after star formation, an approximation justified by the short lifetimes of massive stars relative to the age of the Galactic disk \citep[e.g.,][]{Maeder1989, Padovani1993, Hurley2000}.
We clarify that \citet{Weinberg2024} used the \citet{Magg2022} solar composition, but their calculations ultimately constrain the scale of stellar yields relative to the solar abundance as opposed to the absolute yield.

% \begin{itemize}

%     \item We use O as the representative metal in this work, since it is commonly measured in the gas-phase.
%     Prompt production by massive stars makes O production simple in that it is instantaneous.
%     Previous versions of these models have made predictions for other metals, like production of Fe by Type Ia SNe.

% \end{itemize}

\subsubsection{Helium}
\label{sec:gce:yields:helium}

\citet{Weller2025} investigated which prescriptions for $^4$He yields reproduce observed trends between $^4$He and O abundances.
Based on their recommendations, we use a linear relation between the total $^4$He yield, the O yield, and the solar abundances of O and He (see their equation 12).
Following \citet{Cooke2022}, we use the yields of $^3$He and $^4$He from AGB stars calculated by \citet{Lagarde2011, Lagarde2012}.
We determine population-averaged yields of $^3$He and $^4$He by calculating the total mass released to the ISM after a stellar population has evolved for 10 Gyr.
We adopt a massive star $^4$He yield that makes up the difference between the AGB yield and the total inferred by \citet{Weller2025}.
Weller et al. (2025, in preparation) make further recommendations for $^3$He yields based on $^3$He abundance measurements available in the literature.
The massive star yield of $^3$He is subdominant, so production of this isotope is driven by AGB star yields anyway.
Following \citet{Cooke2022}, we use the non-rotating massive star $^3$He yields from \citet{Limongi2018}.
\par
We retain the primordial $^3$He and $^4$He abundances from \citet{Cooke2022}.
Our primordial helium abundance (by mass) is $Y_\text{P} = 0.24721$ based on \citet{Pitrou2021}, and our primordial helium isotope ratio (by number) is $\heratio\ = 1.257 \times 10^{-4}$.
For both \heratio\ and D/H, adjusting the primordial composition simply shifts the predicted isotope ratios up or down uniformly across the Galaxy, so the exact choice is not particularly important for our purposes.
However, we note that the primordial \heratio\ has never been directly measured, while the primordial D/H has (see discussion below and in Section \ref{sec:discussion:primordial-he-ratio}).

% \begin{itemize}

%     \item We use the prescription for $^4$He nucleosynthesis from \citet{Weller2025} and the prescription for $^3$He from {\color{red} Weller et al. (2025, in preparationn)}.
%     \citet{Weller2025} provide a formula for the total, population-averaged $^4$He yield in terms of the solar O abundance and the O yield.
%     We use the yields predicted by {\color{red} Lagarde et al. (2011?)} for asymptotic giant branch (AGB) star production of $^3$He and $^4$He and {\color{red} Limongi \& Chieffi (2018)} for massive star yields.
    
%     \item Following the recommendations of \citet{Weller2025}, we hold this AGB star yield fixed and scale the massive star yield of $^4$He up and down to match their recommended total yield.
%     Following {\color{red} Weller et al. (2025, in preparation)}, we hold the yields from both massive stars and AGB stars fixed according to the theoretical predictions.
%     The massive star yields of $^3$He are subdominant, while the AGB yields of $^4$He are subdominant.

% \end{itemize}

\subsubsection{Deuterium}

The evolution of D is arguably the simplest of all atomic nuclei.
Protostars are fully convective, which eventually draws all deuterium nuclei into layers hot enough to fuse them into helium \citep{Bodenheimer1966, Mazzitelli1980}.
Consequently, all D nuclei that are incorporated into stars should be destroyed.
We therefore set all D yields to zero and shut off the return of D to the ISM from stellar envelopes in \vice.
The notion that stars both synthesize metals and destroy deuterium leads to a direct one-to-one relationship between metallicity and D/H \citep{Weinberg2017a, vandeVoort2018}.
We use a primordial D abundance of $(\text{D/H})_p = 2.53 \times 10^{-5}$ from \citet{Cooke2018a} in this paper.

% \begin{itemize}

%     \item D has arguably the simplest nucleosynthesis of all element isotopes in that it is believed to be destroyed by stars.
%     Protostars are fully convective, which eventually draws all deuterium nuclei into layers hot enough to fuse them into helium \citep{Bodenheimer1966, Mazzitelli1980}.
%     Therefore, we simply set the yield to zero and shut off ``recycling'' from stellar envelopes returning to the ISM.
%     D is supplied to the ISM through accretion but depletes thereafter.

% \end{itemize}

% \vspace{1.5cm} % prevents this section header from being at the bottom of a page.
\subsection{Ejection versus Radial Gas Flows}
\label{sec:gce:ejection-vs-radialflows}

\begin{figure*}
\centering
\includegraphics[scale = 0.9]{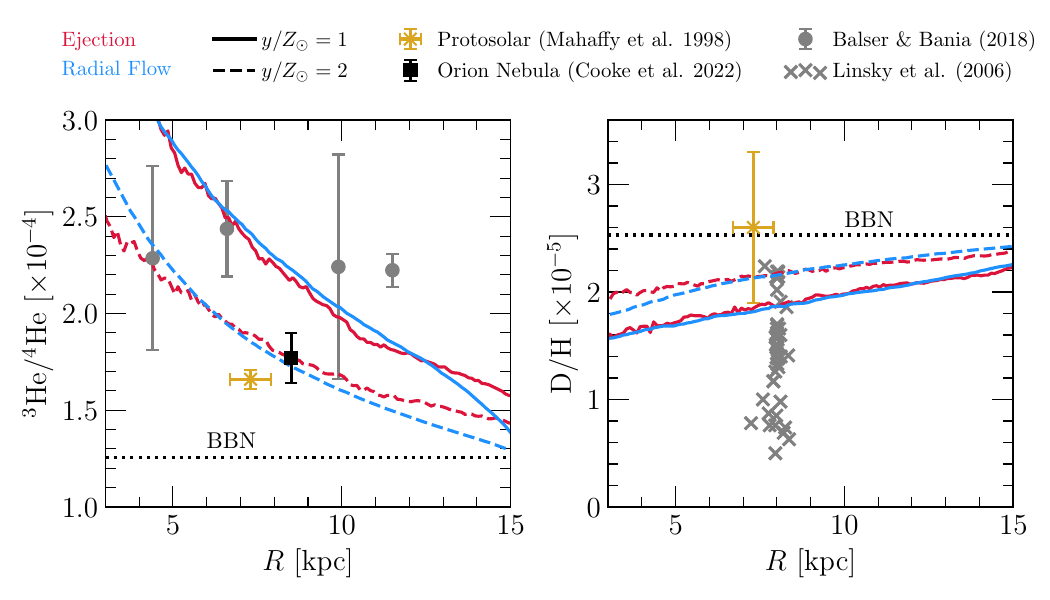}
\includegraphics[scale = 0.8]{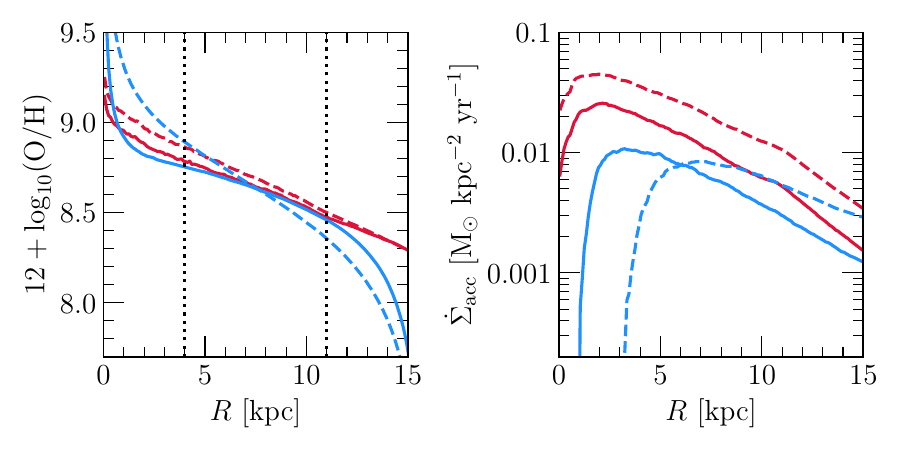}
\caption{
Present day radial profiles of the $\heratio$ and D/H ratios (top), the O abundance (bottom left), and the surface density of gravitational accretion (bottom right).
Top panels show measurements of the non-metal isotope ratios available in the literature (see discussion in Section \ref{sec:data}).
% The black X at $R = 10$ kpc in the top-right shows the median D/H in the \citet{Linsky2006} sample.
The lower left panel highlights the $R \sim 4 - 11$ kpc range as the region of the MW where these data reside.
\textbf{Summary}: Overall, our GCE models have similar profiles in metallicity, but different yield scales lead to different normalizations in the D/H and \heratio\ profiles.
}
\label{fig:present-day-profiles}
\end{figure*}

% \begin{itemize}

%     \item We use two GCE models incorporating ejection from the ISM and two models incorporating radial gas flows within the disk.

%     We provide a brief summary of the key model components in the ejection and radial gas flow scenarios but otherwise leave the details to \citet{Johnson2025} and {\color{red} Johnson (2025)}.

% \end{itemize}

\subsubsection{Ejection-Driven Models}
\label{sec:gce:ejection-vs-radialflows:ejection}

Our models adopted from \citet{Johnson2025} incorporate ejection from the ISM due to feedback.
The mass-loading factor, $\eta \equiv \dot\Sigma_\text{wind} / \dot\Sigma_\star$, describes the mass ejected due to Galactic winds per unit mass of star formation (see discussion in, e.g., the reviews by \citealt{Veilleux2020} and \citealt{Thompson2024}).
These models use an exponential scaling with Galactocentric radius,
\begin{equation}
\eta = \eta_\odot e^{(R - R_\odot) / r_\eta},
\end{equation}
where $R_\odot = 8$ kpc is the location of the Sun \citep[e.g.,][]{GRAVITYCollaboration2019}.
The observed metallicity gradient in the Galactic disk reflects the observed gradient if $r_\eta = 7$ kpc, with $\eta_\odot = 0.4$ for $y / Z_\odot = 1$ and $\eta_\odot = 1.4$ for $y / Z_\odot = 2$ \citep{Johnson2025}.

% \begin{itemize}

%     \item Models from \citet{Johnson2025} incorporate ejection from the ISM due to feedback.
%     Therein we showed that a chemical equilibrium for the MW arises if the mass-loading factor scales with radius quasi-exponentially (i.e., $\eta \propto e^{R}$.
%     $\eta_\odot = 0.4$ and $\eta_\odot = 1.4$ at $R = R_\odot$ sets the normalization for the exponential.
%     Used for $y / Z_\odot = 1$ and $y / Z_\odot = 2$ yield scales, respectively.

% \end{itemize}

\subsubsection{Radial Gas Flow-Driven Models}

Our models in \citet{Johnson2025-solo} use the inward flow of gas toward the Galactic center \citep[e.g.,][]{Lacey1985, Portinari2000, Spitoni2011, Bilitewski2012, Pezzulli2016}.
We examined several prescriptions for the radial velocity of the ISM, finding that equilibrium chemical evolution arises when the flow speed is relatively constant in both radius and time.
Based on our analytic models therein, we use $v_{r,g} = -0.5$ km/s and $-1.8$ km/s as the velocities associated with the $y / Z_\odot = 1$ and $2$ yield scales, respectively.
These models include a weak, centrally concentrated ejection process in order to counteract the pile-up of material in the inner Galaxy \citep[see discussion in][]{Johnson2025-solo}.
This prescription satisfies only this criterion, with $\eta_\odot = 0.2$ and $r_\eta = -5$ kpc, which leaves ejection sufficiently weak across much of the disk that the radial flow is a much more dominant effect on chemical enrichment.

% \red{Weak outflow included to avoid catastrophic pile-up in the inner regions, but evolution at $R \gtrsim 3$ kpc is dominated by the radial flow.}

% \begin{itemize}

%     \item Models from {\color{red} Johnson (2025)} omit ejection from the ISM and use radial gas flows instead.
%     Therein, we showed that the equilibrium scenario arises if the radial velocity in the ISM, $v_{r,g}$, is relatively constant in both Galactocentric radius and time.
%     We use $v_{r,g} = -0.5$ and $-1.8$ km/s as the velocities associated with the $y / Z_\odot = 1$ and $2$ yield scales, respectively.

% \end{itemize}

% \section{Model Predictions}

\section{The Degeneracy}
\label{sec:results}

\begin{figure*}
\centering
\includegraphics[scale = 1]{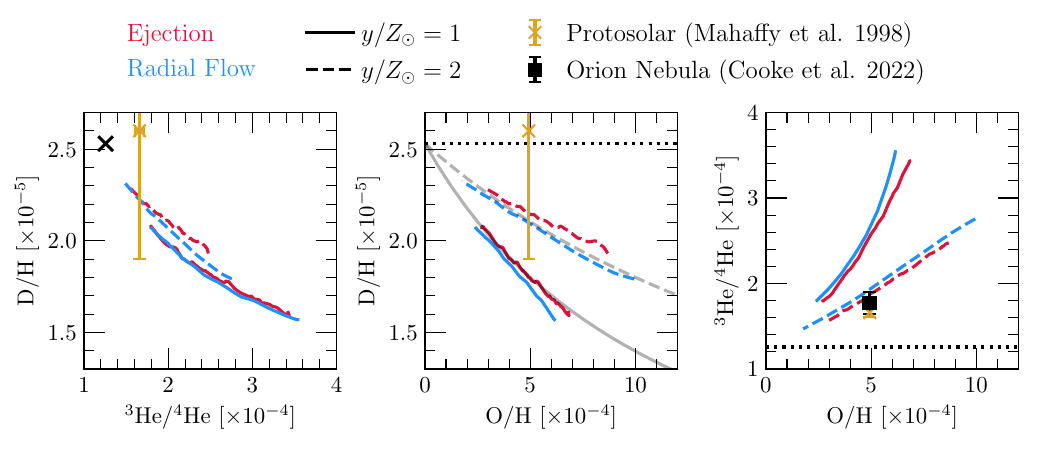}
\caption{
The relationship between O/H, D/H, and \heratio.
Each panel shows the relationship between a pair of two quantities traced by the present-day ISM between $R = 3$ and $12$ kpc.
Lines are colored and styled based on the GCE model with yield scales $y / Z_\odot = 1$ or $2$ (solid or dashed) and adopting ejection or radial gas flows (red or blue).
Our adopted primordial isotope ratios (see Section \ref{sec:gce:yields}) are shown as a black $\times$ in the left panel and black dotted lines in the middle and right panels.
At present, the Sun is the only astrophysical system with a measurement in all three of these panels, though the measurement corresponds to the protosolar composition as opposed to the present day \citep{Mahaffy1998}.
Grey lines in the middle panel show the D/H-O/H relation computed by \citet{Weinberg2017a} for $y/Z_\odot = 1$ (solid) and $y / Z_\odot = 2$ (dashed).
The black square in the right panel shows our measurement of the \heratio\ ratio in the Orion Nebula \citep{Cooke2022} assuming solar metallicity \citep[e.g.,][]{D'Orazi2009}.
\textbf{Summary}: This three-way relationship between O/H, D/H, and \heratio\ can provide measurements of both the scale of stellar yields and the primordial D/H and \heratio\ ratios simultaneously.
}
\label{fig:isotopes-vs-isotopes}
\end{figure*}

% \subsection{The Yield-Ejection-Radial Velocity Degeneracy}
% \label{sec:results:degeneracy}

% We start by describing the three-way degeneracy between stellar yields, the mass-loading factor, and radial flow velocities.
Figure \ref{fig:degeneracy} shows the evolution in the O abundance at $R = 8$ kpc over time in our GCE models at the two scales of stellar yields that we consider.
The predicted evolution traces the yield scale much more closely than the choice of ejection or radial gas flows.
The present-day ISM metallicities of each model are not distinguishable empirically, indicating that present-day gas-phase abundance measurements cannot confidently discriminate between yield scales.
Others have quantified the degeneracy between yields and ejection rates in the past \citep[e.g.,][]{Hartwick1976, Cooke2022, Johnson2023c, Sandford2024}.
Here, we find a similar relationship between the yield and the ISM radial velocity.
This comparison also indicates that it would be difficult to empirically distinguish the effects of ejection from radial gas flows.
\citet{Johnson2025-solo} discusses this topic in the context of metal abundances.
\par
In light of Figure \ref{fig:degeneracy}, one possible means with which to break this degeneracy is by comparing the metal abundances in stars with their ages.
Distinguishing between the $y / Z_\odot = 1$ and $y / Z_\odot = 2$ scenarios requires an abundance precision of $\ll 0.2$ dex in $\sim$$10$ Gyr old populations.
Modern spectroscopic surveys achieve statistical uncertainties in $\alpha$-capture and iron-peak elements at this level (e.g., APOGEE, \citealt{Majewski2017}; SDSS-V Milky Way Mapper, \citealt{Meszaros2025}; GALAH, \citealt{Buder2025}).
However, abundances vary systematically between suveys at the $\sim$$0.1$ dex level \citep[e.g.,][]{Griffith2022, Hegedus2023}.
Stellar age measurements are also substantially challenging, especially for old populations (see the reviews by, e.g., \citealt{Soderblom2010} and \mbox{\citealt{Chaplin2013}}).
In \citet{Johnson2025}, we discussed how the strikingly flat nature of the observed age-metallicity relation tentatively favors the $y / Z_\odot = 2$ scale.
However, a $y / Z_\odot = 1$ model can achieve good agreement with the data by incorporating either pre-enriched accretion from the CGM or metallicity-dependent supernova rates (see discussion in Section \ref{sec:intro}).
Breaking this yield-ejection-velocity degeneracy with this approach is therefore not without substantial nuance in model parameterizations, compounding the issues from measurement uncertainties.

\begin{figure*}
\centering
\includegraphics[scale = 0.5]{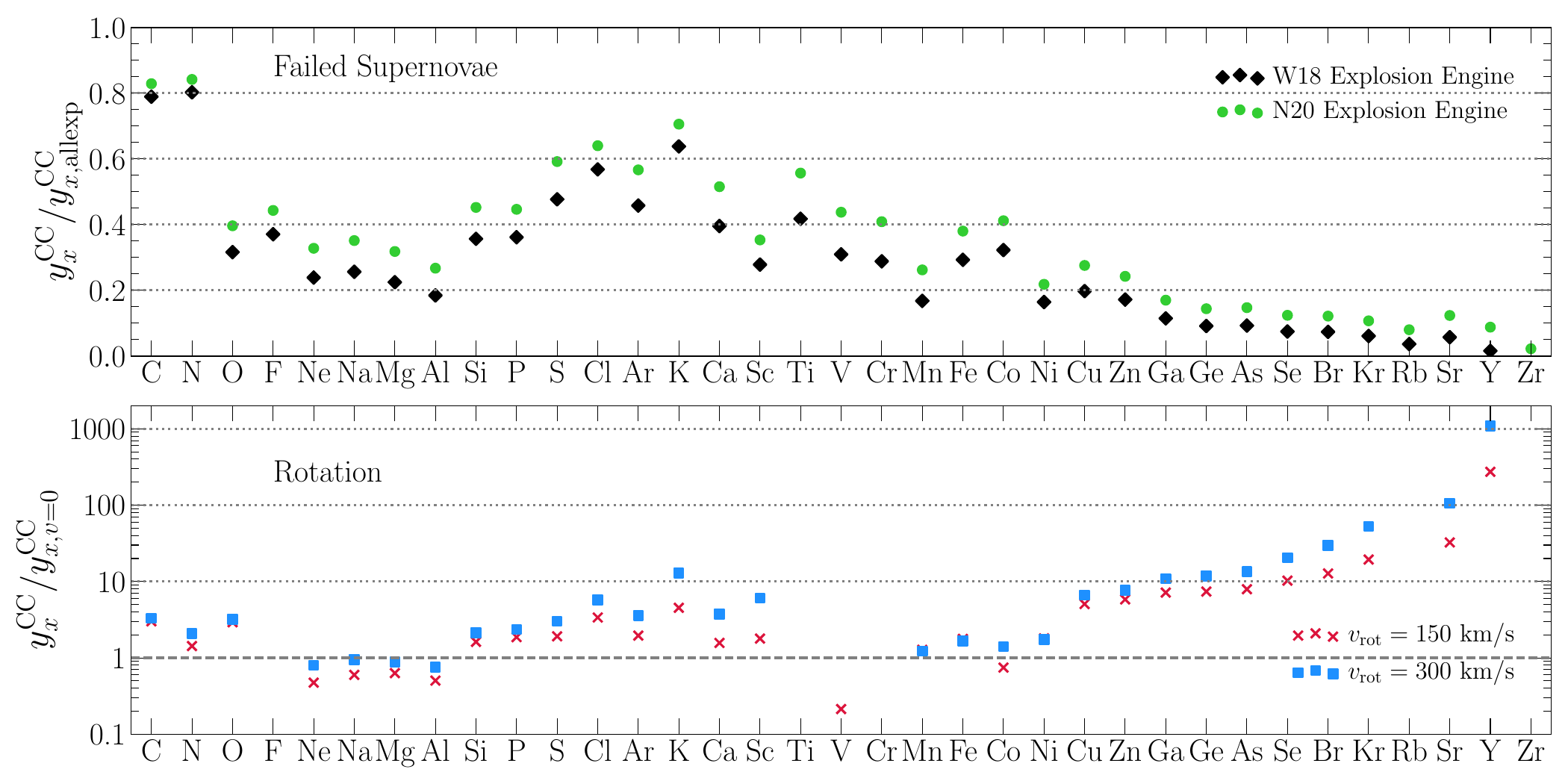}
\caption{
% {\color{red} Put ``failed supernovae'' and ``rotation'' in text on the panels.
% What happened to the other point for V and both points for Cr and Zr in the bottom panel?}
The effects of failed supernovae (top) and rotation (bottom) on population-averaged massive star yields, computed with \vice's \texttt{vice.yields.ccsne.fractional} function using stellar model predictions available in the literature (see discussion in Section \ref{sec:discussion:yield-scale}).
\textbf{Top}: The yields of various elements under the W18 (black diamonds) and N20 (green circles) explosion mechanisms from \citet{Sukhbold2016}, relative to the case where all massive stars explode as a CCSN at the ends of their lives \citep{Griffith2021}.
\textbf{Bottom}: The yields of the same elements assuming initial rotational velocities of $v_\text{rot} = 150$ km/s (red $\times$'s) and $v_\text{rot} = 300$ km/s (blue squares), relative to the non-rotating case ($v_\text{rot} = 0$).
\textbf{Summary}: Failed supernovae and rotation each individually introduce factor of a $\gtrsim$few uncertainties in the predicted yields of most metals, so an empirical benchmark would provide an important test for stellar models.
}
\label{fig:yields}
\end{figure*}

Figure \ref{fig:present-day-profiles} demonstrates how this degeneracy can be broken using the \heratio\ and the D/H ratios in the ISM.
By construction, each model predicts similar O abundance profiles in the $R \approx 4 - 11$ kpc range, where our models are most applicable to the MW (see discussion in, e.g., \citealt{Johnson2021}).
However, the \heratio\ and D/H profiles follow noticeably different normalizations.
With high yields, the high rates of ejection are accompanied by high rates of accretion, which introduces metal-poor gas to the local ISM.
A similar effect arises with radial gas flows, with the main difference being that the more metal-poor material comes from larger radii in the Galactic disk as opposed to the CGM.
By definition, this more metal-poor gas mixing with the local ISM is closer to the primordial composition.
D/H and \heratio\ both reflect these effects similarly.
We note that the radial flow model predicts lower accretion rates overall than the ejection model.
This difference arises because ejection necessitates additional accretion in order to reach the stellar mass of the MW, whereas radial gas flows are not allowed to change the total mass of the ISM \citep{Johnson2025-solo}.
\par
% \heratio\ is more sensitive to the scale of stellar yields than the choice of ejection or radial gas flows.
% This outcome can be understood by considering the differences in accretion histories between models.
% Higher rates of ejection are accompanied by higher rates of accretion in order to fulfill the stellar mass budget of the MW.
% Consequently, a portion of the high \heratio\ ISM is replaced with gas at the primordial isotope ratio (see also discussion in \citealt{Cooke2022}).
% Similar effects arise with radial gas flows because the inflowing matter is always at lower \heratio\ than the local ISM in any region.
% \par
% The D/H abundance is more sensitive to the choice of ejection or radial gas flows than the scale of yields.
% This difference can be understood by considering the differences in accretion histories.
% Ejection-driven models require considerably more accretion to assemble the Galactic disk than flow-driven models.
% Higher rates of ejection are accompanied by higher rates of accretion in order to fulfill the mass budget of the disk.
% Radial flows, on other hand, never alter the total mass of the Galactic disk due to mass conservation \red{(Johnson 2025)}.
\par
In line with our investigation of age-metallicity trends in \citet{Johnson2025}, the helium isotope ratio of the Orion Nebula seems to favor the $y / Z_\odot = 2$ scale of stellar yields.
The protosolar \heratio\ \citep{Mahaffy1998} is slightly below the model in the local ISM, but we have verified that the two are consistent in the snapshot 4.6 Gyr ago.
The protosolar D/H, however, is not measured with sufficient precision for our purposes, nor are the \heratio\ measurements in the radio \citep{Balser2018}.
The \citet{Linsky2006} D/H sample exhibits substantial scatter, which they interpreted as a consequence of deuterium depletion onto dust grains.
If this depletion dominates the scatter, then the true D/H of the ISM should correspond to the upper envelope of the distribution at $\gtrsim$$2 \times 10^{-5}$ \citep{Prodanovic2010}.
Both scales of stellar yields are broadly consistent with this benchmark.
\par
Figure \ref{fig:isotopes-vs-isotopes} shows the three-way relationship between O/H, D/H, and \heratio\ in the present-day ISM.
Each curve shows the corresponding pair of quantities in the ISM between $R = 3$ and $12$ kpc.
Currently, the Sun is the only system with a measurement in all three of these panels.
The left panel shows that D/H and \heratio\ follow a tight relation regardless of the choice of ejection or radial flows.
This outcome arises because deuterium burning leaves behind $^3$He as a byproduct through the d(p,$\gamma$)$^3$He reaction.
The middle panel of Figure \ref{fig:isotopes-vs-isotopes} indicates that ejection and radial gas flows lead to similar D/H-O/H relations.
The normalization is most directly sensitive to the scale of stellar yields for the same reasons discussed above in the context of Figure \ref{fig:present-day-profiles}.
\citet{Weinberg2017a} found this connection between D/H and $y/Z_\odot$ in one-zone models using ejection.
Figure \ref{fig:isotopes-vs-isotopes} indicates that our models are in excellent agreement with the D/H-O/H relation that they calculated, even with radial gas flows.
\citet{vandeVoort2018} found similar results in hydrodynamic simulations.
The right panel of Figure \ref{fig:isotopes-vs-isotopes} indicates that the \heratio\ ratio follows the same relationship with O/H as D/H.
This three-way trend could be useful for sharpening the precision of measurements of the primordial \heratio\ ratio (see discussion in Section \ref{sec:discussion:primordial-he-ratio} below).

\section{Discussion}
\label{sec:discussion}

\subsection{The Scale of Stellar Yields}
\label{sec:discussion:yield-scale}

In this section, we highlight prior work showcasing the potential impact of determining the scale of stellar yields empirically.
We compute population-averaged yields from massive stars for all elements between carbon and zirconium using tables of massive star models available in the literature.
We use \vice's \texttt{vice.yields.ccsne.fractional} function, which computes the yield according to
\begin{equation}
y_x^\text{CC} = \ddfrac{
    \int_8^u \left(
    E(m) m_x + w_x - Z_x (m - m_\text{rem})
    \right)
    \frac{dN}{dm} dm
}{
    \int_l^u m \frac{dN}{dm} dm
},
\label{eq:yxcc}
\end{equation}
where $m_x$ is the explosive yield of the element $x$ from a star of initial mass $m$, $E(m)$ is the fraction of stars at that mass that explode, $w_x$ is the yield of $x$ that is released through stellar winds, and $m_\text{rem}$ is the remnant mass.
The corrective factor $Z_x (m - m_\text{rem})$ converts from a gross yield to a net yield by subtracting off the mass in the ejecta that was present when the stars formed.
The denominator simply normalizes by the total mass of stars formed across the IMF.
\par
The top panel of Figure \ref{fig:yields} shows the effects of failed supernovae on the population-averaged yield, which affects $E(m)$ in Equation \ref{eq:yxcc} above.
We use the yields computed by \citet{Sukhbold2016}, wherein the outcome of explosion or non-explosion is determined self-consistently through the neutrino mechanism.
We focus on their ``W18'' and ``N20'' explosion models and normalize the yields by the results with $E(m) = 1$, which \citet{Griffith2021} computed by forcing explosions where the W18 model otherwise collapsed.
The bottom panel of Figure \ref{fig:yields} isolates the effects of massive star rotation by solving Equation \ref{eq:yxcc} for stellar models with different initial angular momenta.
We use the tables provided by \citet{Limongi2018} at $v_\text{rot} = 150$ and $300$ km/s, normalizing by the predictions with $v_\text{rot} = 0$.
\par
Broadly, failed supernovae and rotation affect yields by factors of a $\sim$few for most elements.
C and N are minimally affected by black hole formation, since a large portion of their yield emerges from massive star winds (see discussion in \citealt{Griffith2021}).
For $\alpha$-capture and iron-peak elements, failed supernovae reduce yields by factors of $\sim$$2-3$.
The effects of rotation are at a similar scale for these elements.
Some elements (e.g., Ti, Cr, Rb) have negative net yields with $v_\text{rot} = 0$ but positive yields with $v_\text{rot} > 0$ and therefore do not show up in the bottom panel of Figure \ref{fig:yields}.
For the heaviest elements shown here, which approach the first s-process peak at the Sr-Y-Zr group, rotation and failed supernovae effect yields by an order of magnitude or more.
These effects underscore the notion that empirical benchmarks of the scale of stellar yields could be useful in pinning down uncertain processes, of which rotation and failed supernovae are only two examples (see discussion in Section \ref{sec:intro}).

\subsection{The Primordial Helium Isotope Ratio}
\label{sec:discussion:primordial-he-ratio}

\begin{figure}
\centering
\includegraphics[scale = 1]{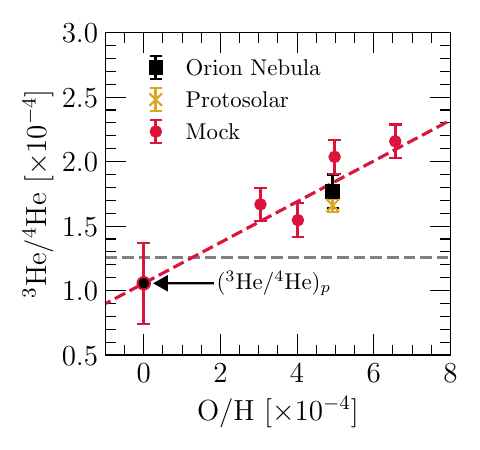}
\caption{
An inference of the primordial $\heratio$ ratio, $(\heratio)_p$, using available measurements from the literature combined with mock data from our GCE model using ejection with the $y / Z_\odot = 2$ yield scale (see discussion in Section \ref{sec:gce}).
The black square and gold $\times$ symbol indicate measurements of $\heratio$ in the Orion Nebula by \citet{Cooke2022} and in Jupiter's atmosphere by \citet{Mahaffy1998}, respectively.
Red circles show mock measurements drawn from the GCE model at the present day at Galactocentric radii of $R = 5$, 7, 9, and 11 kpc, offset by measurement uncertainties of $\sigma([\text{O/H}]) = 0.02$ and $\sigma(^3\text{He}/^4\text{He}) = 0.13 \times 10^{-4}$ (see discussion in Section \ref{sec:discussion:primordial-he-ratio}).
The red dashed line shows the line of best-fit to these data, with the intercept at O/H $ = 0$ marked as the inferred primordial isotope rate of 
$(^3\text{He}/^4\text{He})_p = (1.06 \pm 0.32) \times 10^{-4}$.
The grey dashed line marks the value of $(\heratio)_p$ used as input to our GCE models.
\textbf{Summary}: Measurements of the $\heratio$ ratio along $\sim$$4$ additional sightlines within distances of $\lesssim$$3$ kpc, each achieving similar precision as \citet{Cooke2022}, would provide a measurement of $(\heratio)_p$ with $\sim$$30$\% precision.
}
\label{fig:primordial-he-ratio-inference}
\end{figure}

In this section, we demonstrate that mapping \heratio\ in the ISM across the MW would enable the first measurement of the primordial helium isotope ratio, $(\heratio)_p$.
To this end, we construct a mock smaple of \heratio\ measurements using our ejection-driven GCE model with the $y / Z_\odot = 2$ yield scale.
This model most clearly reproduces the observed trends in metallicity with stellar age across the Galactic disk (see discussion in \citealt{Johnson2025} and \citealt{Johnson2025-solo}).
We start with the present-day ISM composition at Galactocentric radii of $R = 5$, $7$, $9$, and $11$ kpc.
We randomly sample offsets from the predicted O abundances to mimic a measurement uncertainty of $\sigma([\text{O/H}]) = 0.02$, which is typical of modern high-resolution spectroscopic surveys \citep[e.g.,][]{Buder2025, Meszaros2025}.
We use an uncertainty of $\sigma(\heratio) = 0.13$, which is the precision reached by \citet{Cooke2022} for the Orion Nebula.
Measurements in more distant \hii\ regions may carry higher uncertainties, but we use this fiducial value for this simple forecast.
\par
Figure \ref{fig:primordial-he-ratio-inference} shows the resulting mock sample alongside the existing measurements by \citet{Cooke2022} and in Jupiter's atmosphere by \citet{Mahaffy1998}.
We assume solar O/H for the Orion Nebula, consistent with observations \citep{D'Orazi2009}.
Following previous estimates of the primordial \textit{elemental} abundance of He \citep[e.g.,][]{Peimbert1974, Izotov2014, Peimbert2016, Fernandez2019, Valerdi2019, Hsyu2020, Aver2021}, the primordial isotope ratio can be determined by extrapolating this trend to O/H $= 0$.
We perform a linear regression to these six data points, finding a slope of $0.158 \pm 0.065$ and an intercept of $(\heratio)_p = (1.06 \pm 0.32) \times 10^{-4}$.
These uncertainties correspond to $\sim$40\% precision of the slope of the trend and $\sim$30\% precision of the primordial helium isotope ratio.
We explored variations in the random number seed used to draw the mock sample, finding uncertainties in $(\heratio)_p$ ranging from $\sim$15\% to $\sim$40\% precision.
We select this particular random subsample because it is typical but slightly conservative compared to other random draws.
In general, measurements across a broad range of Galactocentric radii lead to a higher forecasted precision in $(\heratio)_p$.
This outcome arises because sampling a range of Galactocentric radii also samples a range of metallicity due to the radial abundance gradient (see Figure \ref{fig:present-day-profiles}).
In detail, the \heratio-O/H relation is slightly non-linear, more noticeably so in the $y / Z_\odot = 1$ scale than $y / Z_\odot = 2$ (see Figure \ref{fig:isotopes-vs-isotopes}).
The degree of non-linearity is subtle, so a linear relation should suffice for the foreseeable future, until the sample size and measurement precision of \heratio\ improve substantially.

\subsection{Measurements}
\label{sec:discussion:measurements}

% \red{Pollux survey -- part of Habitable Worlds Observatory.
% Potentially useful -- original plan was to do far UV all the way to IR, could do \heratio\ and D/H simultaneously.
% Seems like the wavelength range has been revised down in planning.
% Latest report from \citet{Muslimov2024} indicates possibility that Pollux could cover the $\lambda = 100 - 120$ nm range, which would be useful but not ideal for Lyman series transitions.
% Could get Lyman-alpha blueshifted into this range or Lyman-beta redshift away from the $100$ nm break?
% }

\subsubsection{Helium}
\label{sec:discussion:measurements:he}

The most reliable method for measuring the helium isotope ratio of the ISM follows \citet{Cooke2022}.
This approach uses the light from a background star and measures the column density of helium in the metastable nuclear excited state, \heistar, in absorption due to a foreground screen of dense and highly ionized gas.
Initially described in \citet{Cooke2015}, this method compares the line profiles of multiple \heistar\ transitions.
In principle, differences in line profiles provide an unambiguous measurement of \heratio\ because the isotope shift is different for all \heistar\ transitions.
The ionization potentials of $^3$He and $^4$He are also nearly identical, eliminating the need for an ionization correction.
In \citet{Cooke2022}, we used the transitions at $\lambda3188$ \AA, $\lambda3889$ \AA, and $\lambda1.0833\ \mu$m.
The optical transitions have isotope shifts of only $\sim$$10$ km/s, so their line profiles are dominated by $^4$\heistar.
The $^3$\heistar\ column density then follows from the $\lambda1.0833\ \mu$m line, where the isotope shift is $\sim$$35$ km/s.
% One could also use the optical transition at $\lambda6679$ \AA, but the isotope shift is smaller ($\sim$22.5 km/s).
% This difference necessitates higher signal-to-noise than the $\lambda1.0833\ \mu$m line to make a similar measurement.
\par
An important challenge with this approach is that metastable helium is rare.
Historically, efforts to detect \heistar\ have often yielded upper limits (see discussion in \citealt{Indriolo2009}).
Successful detections have been reported along sightlines toward hot stars in the Orion Nebula \citep{Wilson1937, O'Dell1993, Oudmaijer1997}, in NGC 6611 \citep{Evans2005}, and toward $\zeta$ Ophiuchus \citep{Galazutdinov2012}.
Helium nuclei only exist in the metastable state when He$^+$ recaptures an electron that has the same spin as the one already present in the He$^+$ ion.
Therefore, the observed absorption likely arises at the edge of the He \textsc{ii} ionization region.
Wolf-Rayets, blue supergiants, and O/B stars are therefore potentially useful targets, since they are hot enough to ionize He.
Hot stars also have implicitly clean spectra near the \heistar\ transitions of interest.
In red and yellow supergiants, carbon features appear near the $1.0833\ \mu$m line.
\par
Ground state helium is unfortunately not useful for measuring the isotope ratio.
Available measurements in the radio \citep[][see Figure \ref{fig:present-day-profiles} and discussion in Section \ref{sec:results}]{Balser2018} do not reach sufficient precision for these purposes.
All other helium transitions are in the far-UV, which would require observatories in space and sightlines that are not attenuated by \hi\ absorption in the same gas cloud \citep[see, e.g.,][]{Cooke2018b}.
However, the isotope shifts in the UV are small, of order $\sim$$12$ km/s.
The $\lambda1.0833\ \mu$m line has the largest isotope shift ($\sim$$35$ km/s) of all helium transitions.
Therefore, for the foreseeable future, the best path toward measuring the Galactic distribution of \heratio\ is through the metastable transitions along sightlines toward massive stars.

\subsubsection{Deuterium}
\label{sec:discussion:measurements:d}

D/H is most often measured using Lyman series transitions \citep[e.g.,][]{Rogerson1973, Adams1976}.
% Lower energy transitions like the Balmer series are not sufficiently strong spectral lines to confidently isolate D and H.
Historically, D/H measurements have been made along sightlines toward stars near the Sun (e.g., \citealt{Linsky2006} and references therein) and toward high-redshift quasars or damped Lyman-$\alpha$ systems (see, e.g., the review by \citealt{Cooke2026}).
The former provide measurements of D/H in the local ISM that we have used in this paper (see Figure \ref{fig:present-day-profiles}), while the latter have been used to infer the primordial D/H ratio \citep[e.g.,][]{Cooke2018a, Kislitsyn2024}.
With an isotope shift of $\sim$82 km/s in Lyman series transitions, D \textsc{i} and \hi\ can be reliably distinguished from one another with spectral resolution of $R \gtrsim 25,000$.
D/H measurements therefore do not face the same technical challenges as \heratio\ measurements (see discussion in Section \ref{sec:discussion:measurements:he} above).
However, the Lyman series is sufficiently far into the ultraviolet (UV) that MW measurements require a space-based UV-sensitive echelle spectrograph.
The usefulness of mapping D/H across the MW points to a need for continued investment in UV-sensitive facilities in space.

\section{Conclusions}
\label{sec:conclusions}

Building on previous GCE models, we have demonstrated that metal abundances in the MW disk are subject to a ``source-sink'' degeneracy.
Yields of all metals from stars can be uniformly increased, but similar metallicities arise as long as one proportionally ejects more gas from the ISM \citep{Hartwick1976, Cooke2022, Johnson2023c, Sandford2024}.
In this paper, we have shown that radial gas flows \citep[e.g.,][]{Lacey1985, Bilitewski2012, Johnson2025-solo} have qualitatively similar effects.
High stellar yields result in similar metal abundances as lower yields if the inward flow velocity is proportionally faster (see Figure \ref{fig:degeneracy}).
\par
This yield-ejection-velocity degeneracy can be broken by mapping the \heratio\ or D/H isotope ratios in the ISM across the MW disk (see Figure \ref{fig:present-day-profiles}).
Non-metals have different production and destruction channels than metals.
By tracing different processes, measurements of hydrogen and helium isotopes can break this degeneracy.
Our models predict similar radial profiles in both \heratio\ and D/H when swapping ejection for radial gas flows, indicating that this connection is independent of the assumed set of GCE parameters.
Non-metal isotope ratios are therefore useful empirical diagnostics of the scale of stellar yields, in line with prior work \citep[e.g.,][]{Weinberg2017a, Cooke2022}.
% \heratio\ when swapping ejection for radial gas flows, but differ based on the choice of stellar yield.
% D/H is instead more sensitive to the choice of ejection versus radial gas flows.
% Our ejection-driven models require significantly more accretion to build the Galactic disk than our radial flow models, which introduces more deuterium into the Galaxy.
% We identify the D/H--\heratio\ and \heratio--O/H planes are usefully unambiguous diagnostics of these processes (see Figure \ref{fig:isotopes-vs-isotopes}).
\par
Breaking this degeneracy would provide a useful benchmark for stellar evolution models.
Poorly understood processes introduce uncertainties in metal yields at the factor of a $\gtrsim$few level for most elements (see Figure \ref{fig:yields} and discussion in Section \ref{sec:discussion:yield-scale}).
We have highlighted this result for failed supernovae \citep{Griffith2021} and rotation \citep[][see Figure \ref{fig:yields}]{Limongi2018}, but similar uncertainties are introduced by mass loss \citep[e.g.,][]{Sukhbold2016}, convection \citep[e.g.,][]{Chieffi2001}, and nuclear reaction rates \citep[e.g.,][]{Herwig2004}.
Precise empirical constraints on the total metal yield from stellar populations would allow stellar evolution and supernova models to be ruled out based on over- or under-production of certain elements.
Given dust depletion of D/H in the ISM \citep{Linsky2006}, \heratio\ might be a better tracer of the effects discussed by \citet{Weinberg2017a}.
For example, measurements of \heratio\ may allow us to test if the MW disk exhibits large metallicity variations due to accretion of low-metallicity gas \citep[e.g.,][]{DeCia2021}.
If \heratio\ varies smoothly across the Galactic disk, that result would confirm the long-standing suggestions that the low D/H along some sightlines arises due to dust depletion.
\par
Mapping \heratio\ across the Galactic disk would also enable the first inference of the primordial helium isotope ratio, $(\heratio)_p$ (see Figure \ref{fig:primordial-he-ratio-inference}).
With four additional \heratio\ measurements within $\sim$3 kpc of the Sun, we project a $\sim$30\% statistical uncertainty in $(\heratio)_p$.
This primordial helium isotope ratio would provide a yet unachieved test of current models of BBN (see discussion in \citealt{Cooke2022}).
The relationship between D/H, O/H, and \heratio\ could also be of potential use in pinning down the composition of the primordial Universe more precisely (see Figure \ref{fig:isotopes-vs-isotopes}).
\par
Ultimately, the yield-ejection-velocity degeneracy underscores the notion that metal abundances alone are fundamentally limited in their information content.
Specifically, metal production on its own does not specify the metallicity scale.
Metallicity is also influenced by the amount of hydrogen that newly produced metals mix with.
Unsurprisingly, this information is encoded in the relative numbers of hydrogen isotopes.
Breaking this degeneracy provides information on the timescales of metallicity growth in galaxies, which is sensitive to the relative balance of accretion, star formation, and ejection (see discussion in, e.g., \citealt{Johnson2025}).
These processes are central components of galaxy evolution models.
Each method of breaking the degeneracy carries its own challenges and sources of systematic uncertainty (see discussion in Section \ref{sec:intro}).
Although the measurements are challenging (see discussion in Section \ref{sec:discussion:measurements}), a detailed inventory of non-metals would provide constraints useful for a range of purposes in astrophysics.
Currently, there are only a handful of available measurements along sightlines in the MW (see discussion in Section \ref{sec:data}).
Even small additions to the current sample would significantly improve the landscape.

\section*{Acknowledgements}

We are grateful to David Weinberg for valuable discussion of this project.
JWJ acknowledges support from a Carnegie Theoretical Astrophysics Center postdoctoral fellowship.
During this work, RJC was supported by a Royal Society University Research Fellowship.
RJC acknowledges support from STFC (ST/X001075/1).
This research has made use of NASA's Astrophysics Data System.

\bibliographystyle{aasjournal}
\bibliography{main}

\end{document}